\documentstyle[aps,prb,epsfig,floats]{revtex}

\begin{document}
\twocolumn[\hsize\textwidth\columnwidth\hsize
\csname @twocolumnfalse\endcsname
\title{Theory for two-photon photoemission: transport and temperature
  effects} 
\author{R.~Knorren$^*$, G.~Bouzerar, and K.~H.~Bennemann}
\address{Institut f\"ur Theoretische Physik, Freie Universit\"at
Berlin, Arnimallee 14, D-14195 Berlin, Germany} 
\date{\today}  
\maketitle

\begin{abstract}
Using a theory which treats on equal footing transport of excited
electrons and electron-phonon scattering, we are able to explain the
temperature dependence of the relaxation time in Cu as recently
observed by Petek, Nagano, and Ogawa.   We show that the unexpected
increase of the relaxation time with temperature results from the
drastic change of the electron motion due to the efficiency of
electron-phonon scattering:  the transport is ballistic at low
temperature and gets diffusive at room  temperature. Finally, our
theory also reproduces  the experimental measurements of the
two-photon photoemission  (2PPE) intensity as a function of the
pump-probe delay.
\end{abstract}
\pacs{72.15.Lh,78.47.+p}
]
\narrowtext

Due to advances in  the ultrashort laser techniques, very efficient
tools to study the dynamics of excited electrons in semiconductors and
metals are available.  One of the key methods is the time-resolved
two-photon photoemission (TR 2PPE) which has the crucial advantage to
allow a direct measurement of the variation of the electron
distribution.\cite{Schmuttenmaer94,Hertel96,Aeschlimann96,Ogawa97a,Aeschlimann97,Cao98}
Recent measurements on Cu have shown an unusual non-monotonic behavior
of the relaxation time as a function of
energy.\cite{Pawlik97,Cao97,Knoesel98,Petek99b}  Furthermore, an
unexpected increase  of the relaxation time with temperature was
reported.\cite{Petek99} We have recently developed a theory for the
dynamics of  excited electrons in metals, which explicitly includes
the effect of secondary
electrons,\cite{Knorren99a,Knorren99b,Knorren00} in contrast to  {\it
ab-initio} calculations of the electronic
lifetime.\cite{Campillo99,Echenique00,Schoene99,Keyling00}   It
consists in solving a Boltzmann-type equation in the random-{\bf k}
approximation.  Within our approach, we were able to reproduce the
peak in the relaxation time at the right energy and a linear shift
with photon frequency, in agreement with experiments. We have also
analyzed in detail the structure and height of the peak and the role
of secondary  electrons, especially focussing on the Auger
contribution and the $d$-hole lifetime.

To explain the observed temperature dependence, in this paper we
extend our model by including effects of  transport of excited
electrons out of the detection region  and electron-phonon
scattering. Note that the temperature dependence of the relaxation
time is striking, since on the  basis of Fermi-liquid theory (FLT) one
would expect only a very small {\it decrease} of  $
\tau(E,T)=a_{0}/[(E-E_{F})^2 + b(k_B T)^2]$.\cite{Quinn62,Pines}
Since $b$ is of oder 1 and  $k_BT$ is much smaller than $E-E_{F}$, the
reduction is insignificant for excited electrons  of energy of order 1
eV.  But if one assumes that electron-phonon collisions are efficient
enough to change the nature of the transport from a regime almost
ballistic at low temperature to a diffusive regime at room
temperature, then one should expect a longer relaxation time at higher
temperature. This is supported by the observation that ballistic
transport strongly reduces the relaxation
time.\cite{Aeschlimann96,Knoesel98,Knorren99b}

Let us now describe the theoretical approach we use.
The temporal variation of the occupation of a level of energy $E$ and
momentum {\bf k} at distance $z$ from the surface is described by a
Boltzmann-type equation: 
\begin{eqnarray}
\frac{\partial f(E,{\bf k},z)}{\partial t} &=& \left[\frac{\partial f}{\partial
  t}\right]_{opt} +
\left[\frac{\partial f}{\partial
  t}\right]_{e-e} \nonumber\\
&+& \left[\frac{\partial f}{\partial t}\right]_{transp} +
  \left[\frac{\partial f}{\partial t}\right]_{e-ph} \;,
\end{eqnarray}
where respectively the first term corresponds to the optical
excitation, the second  one describes the electron-electron
scattering, the third is the ballistic transport and the fourth is the
electron-phonon scattering.  The transport term is given
by\cite{Ziman72}
\begin{equation}
\left[\frac{\partial f}{\partial t}\right]_{transp} = -v_{z}\frac{\partial
  f}{\partial z} \;.
\end{equation}
Here, $v_{z}$ is the $z$-component of the electron velocity. We only
consider transport in the $z$-direction, because the diameter of the
laser spot is much larger than the optical penetration depth.
Detailed expressions for $\left[\frac{\partial  f}{\partial
t}\right]_{e-e}$ in the random-{\bf k} approximation (including
secondary-electron generation) and for  the optical excitation term
$\left[\frac{\partial  f}{\partial t}\right]_{opt}$ are given in
Ref.~\onlinecite{Knorren99b}.  To derive the electron-phonon
scattering term, we first consider the general
expression,\cite{Grimvall81,Allen87} neglecting the transfer of energy
to the lattice. The exchange of energy between the lattice and the hot
electrons starts to get really effective only after
$0.5$~ps,\cite{Suarez95,Fann92b} which is not the regime we consider
here. Also, in noble  metals there is only one acoustic branch (1
atom/unit cell), and an upper bound for the phonon-energy average is
$\hbar \langle \omega_{ph} \rangle \approx k_B T_{D}$, where $T_{D}$
is the Debye temperature. For Cu $\hbar \langle \omega_{ph} \rangle
\approx 20$~meV,\cite{Brorson80} which can be neglected since it  is
small enough compared to the excited-electron energy ($\approx 1$~eV).
On the other hand, large transfer of momentum is allowed, since the
Debye cut-off is of order $k_{D}a \approx \pi$ ($a$ is the lattice
spacing).\cite{Ashcroft} This is a crucial point in our theory,
because only large   momentum transfer will efficiently change the
direction of propagation of the electrons.  Let us then express within
these simplifications,  the electron-phonon scattering contribution in
the random-{\bf k} approximation:
\begin{eqnarray}
\left[\frac{\partial f(E,{\bf k})}{\partial t}\right]_{e-ph}  &=&
-\Gamma \int  
\frac{d\Omega_{\bf k'}}{4 \pi} [1 -f(E,{\bf k'})]  f(E,{\bf k})
\nonumber\\ 
&& + \Gamma \int \frac{d\Omega_{\bf k'}}{4 \pi} f(E, {\bf k'}) [1-
f(E, {\bf k})] \;,
\label{ep}
\end{eqnarray}
where $\Gamma = \frac{2 \pi}{\hbar} |g|^{2} (2\langle n \rangle+1)
\rho(E_{F})$ is the electron-phonon scattering rate.  $\rho(E_{F})$ is
the density of states at the Fermi surface, $\Omega_{\bf k'}$ denotes
the solid angle, and $\langle n \rangle = \left\{\exp\left[\hbar
\langle \omega_{ph}\rangle/(k_B T)\right]-1\right\}^{-1}$ denotes the
thermal average of the phonon occupation.  Additionally we assume that
the coupling function $g$ is constant.  Noticing that $k_B T \gg \hbar
\langle \omega_{ph}\rangle$, we get the well-known formula
\begin{equation}
\Gamma = \frac{2 \pi}{\hbar} \lambda k_B T \;,
\label{gamma}
\end{equation}
where $\lambda =2 |g|^{2} \rho(E_{f})/(\hbar \langle \omega_{ph}
\rangle)$ is   the so-called electron-phonon mass enhancement
factor.\cite{Grimvall81} The last step in evaluating Eq.~(\ref{ep})
is to make the substitutions $f(E, {\bf k}) \rightarrow f(E,v)$,
where $v$ is the $z$-component of the velocity, and $ \int
\frac{d\Omega_{\bf k'}}{4 \pi} [1 -f(E, {\bf k'})] \rightarrow
\frac{1}{N}\sum_{j} [1- f(E,v_j)]$.

The 2PPE intensity is calculated as the convolution of the probe
laser intensity $P(t)$ with the distribution of excited electrons 
in the vicinity of the surface:\cite{Knorren99b}
\begin{eqnarray}
I^{\rm 2PPE}(E,\Delta t) &=& \int_{-\infty}^{\infty} dt\ 
  P(t-\Delta t) \nonumber\\
&& \times \int_0^{\infty}  dz\ e^{-z/\lambda_{\rm esc}} f(E,z,t)
  \;,
\label{2ppe_signal}
\end{eqnarray}
where $\lambda_{\rm esc} = 1.6$~nm is the escape depth taken from
overlayer experiments.\cite{Siegmann94} The effective relaxation time
$\tau$ is extracted from the 2PPE intensity as a function of
pump-probe delay, as in experiments. 

In order to allow a direct comparison between our calculations and the
available experimental data we also use in the calculations a laser
pulse of duration  $12$~fs and energy $3.1$~eV. For  Cu, we use 15~nm
for the optical penetration depth and $v_t=1.8\rm\ nm/fs$ for the
transport velocity.   We chose $\tau_h=35$~fs for the $d$-hole
lifetime, since it was shown in a previous study that   this provides
a good order of magnitude for the height of the peak in the relaxation
time in Cu.\cite{Knorren00} This is also in agreement with two
independent experimental measurements which have suggested a lower
bound of order $25$~fs.\cite{Matzdorf99,Petek99a} We have no other
free parameter, since the parameter $\lambda$ which enters in the
electron-phonon scattering rate $\Gamma$ was measured to be $\lambda
\approx 0.15$.\cite{Grimvall81,Mcdougall95} Note, by using
Eq.~(\ref{gamma}) one can already get an insight of the effect of the
electron-phonon collisions on the transport: at $T=300$~K the average
time between two elastic collisions is $\tau_{e-ph}=1/\Gamma \approx
30$~fs, in agreement with the value extracted from resistivity
measurements ($\tau_{e-ph}=27$~fs).\cite{Kaye} However, at $T=50$~K,
we get $\tau_{e-ph} \approx 170$~fs.  Thus, since an excited electron
at $1$~eV has a relaxation time of order $\tau_e=50$~fs,  one expects
almost no effect on transport due to electron-phonon scattering at
$T=50$~K, because $\tau_{e-ph} \gg \tau_e$. In contrast, at $T=300$~K,
one has $\tau_{e-ph} \le \tau_e$, and thus the phonons will be very
effective in changing the nature of the transport.

\begin{figure}[t]
\begin{center}
\epsfig{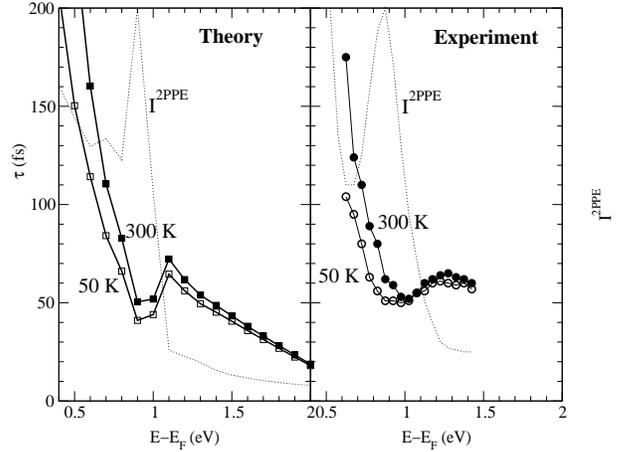} 
\end{center}
\caption{\label{fig1}Temperature dependence of the 
relaxation time $\tau$
(experimental data for Cu(111) from Ref.~\cite{Petek99}).
The dotted line shows the
2PPE spectrum $I^{\rm 2PPE}(\Delta t=0)$ in arbitrary units. 
The pulse duration is $12$~fs and the photon energy 3.1~eV.}
\vspace*{-0cm}
\end{figure}

In Fig.~\ref{fig1} we have plotted both the 2PPE intensity $I^{\rm
2PPE}(\Delta t=0)$ and the relaxation time  $\tau$ at $T=300$~K and
$T=50$~K as a function of energy. We compare our results with the
experimental data of Ref.~\onlinecite{Petek99}.  First, the 2PPE
intensity  in both theoretical and experimental data compare quite
well.  A pronounced peak due to transitions from the $d$ band appears
at $E-E_F=0.9$~eV and is  followed by a sharp threshold at around
1.1~eV. At low  energy, we again observe an increase of the 2PPE
signal. Second, the data for the relaxation time show a surprisingly
good agreement: i) the positions of the peak and the dip in the
relaxation time are identical; ii) the structure is similar, although
the height of the peak is larger in our calculation, which could
indicate that the $d$-hole lifetime $\tau_h$ could be smaller than the
value of 35~fs considered  here; iii) the magnitude of the change due
to the temperature is the same.  For example, at $E=0.6\rm\ eV$ and
$300\rm\  K$ we get for both the experimental and calculated
relaxation time   $\tau_{exp} = 175\rm\  fs$ and  $\tau_{th} = 170\rm\
fs$,  while at $ 50\rm\  K$ we find $\tau_{exp} = 105\rm\  fs $ and
$\tau_{th} = 110\rm\  fs$. This agreement is surprisingly good.  As
expected at sufficiently high energy $ E - E_{F} > 1.5\rm\  eV$ the
relaxation time is almost unaffected, since the excited-electron
lifetime is smaller in both cases than $\tau_{e-ph}$.

\begin{figure}[t]
\begin{center}
\begin{tabular}{cc}
\epsfig{height=6cm,width=4cm,angle=0,file=figure2.eps} 
&
\raisebox{-0.cm}[0pt]{\epsfig{width=4cm,height=6cm,angle=0,file=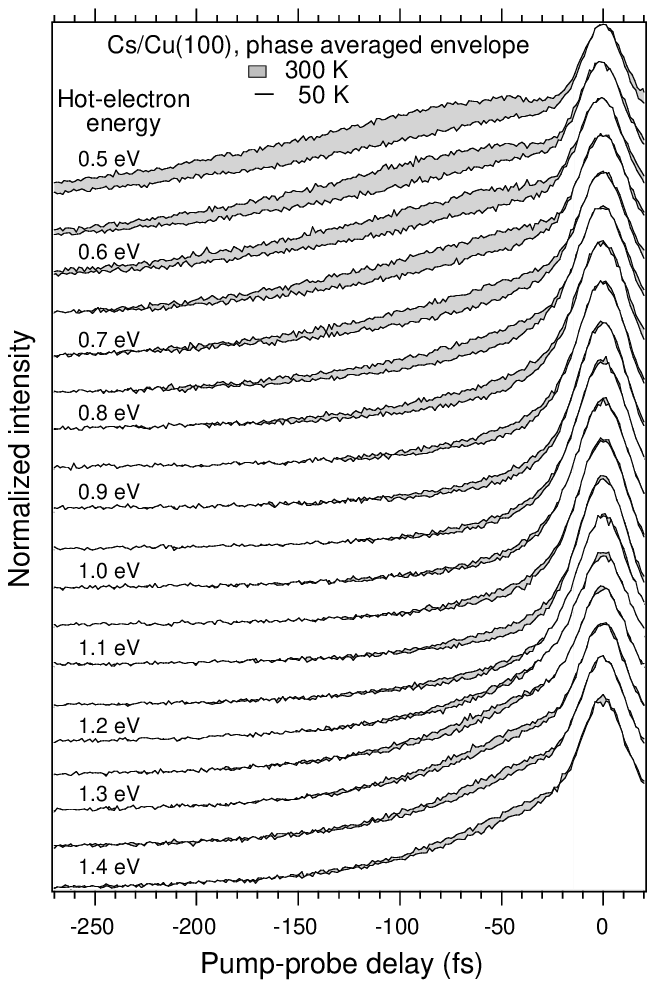}} 
\end{tabular}
\end{center}
\caption{\label{fig2}Temperature dependence of the 
  2PPE intensity for different energies $E-E_F$ as a function of
  pump-probe delay. The data on the right side are experimental
  results from Ref.~\cite{Petek99} The data on the left side are the
  theoretical results. Note the delayed
  rise at low energy for $T=300$~K which leads to the
  larger relaxation time.}
\vspace*{-0cm}
\end{figure}

In Fig.~\ref{fig2} we compare the 2PPE intensity as a function of the
pump-probe delay with the data from Ref.~\onlinecite{Petek99}. One can
observe a strong temperature dependence, especially at low
energy. Again we get very good agreement for both the quantitative and
qualitative aspects. Note that the experimental results include a peak
at $\Delta t=0$, which is absent in the calculations because our
theory does not include coherent  effects.  So one should compare the
data for $\Delta t < -20$~fs. First, at low energy up to 0.7~eV, in
both cases a clear delayed rise is observed at  $T=300$~K.  Second,
the magnitude of the temperature effect agrees surprisingly well with
the experimental data. It is more pronounced at low energy and
decreases with increasing energy. In a very small window of energy
around 1.1~eV (just above the $d$-band threshold), we observe a
reappearance of the delayed peak accompanied by a slight new increase
of the temperature dependence. The increase at 1.1~eV was also noted
in Ref.~\onlinecite{Petek99}, although the delayed rise seems to be
absent.  It is also interesting to remark that such a rise could be
observed for the first time, since a sufficiently short laser pulse
was used. This rise is in fact a signature of the presence of
secondary electrons.

\begin{figure}[t]
\begin{center}
\epsfig{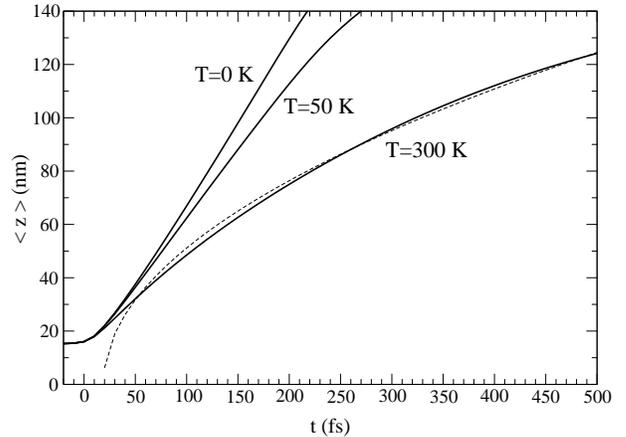} 
\end{center}
\caption{\label{fig3}Average distance $\langle z \rangle$ of excited
  electrons from the 
  surface after laser excitation. Note that
  at $T=0$ and 50~K the transport is ballistic, while at $T=300$~K it is
  diffusive due to elastic electron-phonon collisions. The dashed line
  is a fit using $\langle z \rangle =\sqrt{D(t-t_0)}$.} 
\vspace*{-0cm}
\end{figure}

To  illustrate the reason why the effect of temperature is so strong,
let us  analyze the motion of the excited-electron distribution as a
function of time. As a measure for the penetration of electrons into
the bulk, we define the average distance  from the surface, $\langle
z(t) \rangle$:
\begin{equation}
\langle z(t) \rangle = \frac{\int_0^{\infty} dz z
  N(z,t)}{\int_0^{\infty} dz N(z,t)} \;, 
\end{equation}
where $N(z,t) = \int_0^{\infty} dE \rho(E) f(E,z,t)$ is the average
number of excited electrons at distance $z$ and time $t$.  In
Fig.~\ref{fig3}, we have plotted $\langle z(t) \rangle$ at different
temperatures, $T=0$, 50, and 300~K.   Clearly we observe that at $T=0$
and 50~K, the motion of the excited electrons is ballistic. At
$T=50$~K,  a small deviation from the linear behavior appears  around
$t\approx 200$~fs, which is of the order of magnitude of
$\tau_{e-ph}=170$~fs. We get for the average velocity $\Delta
\langle z \rangle / \Delta t \approx v_t/2$, where the factor 1/2 can be
understood easily by considering the average of the velocity in $z$
direction. However, the nature of the motion has drastically changed
at $T=300$~K: the motion is now diffusive. We illustrate this  by
fitting the data with $\langle z \rangle = \sqrt{D(t-t_0)}$, which is
expected in the case of diffusive motion.  An offset $t_0$ is
introduced in order to take into account the finite duration of the
laser pulse generating excited electrons. We get $D=32\rm\ nm^2/fs$,
which agrees very well  with the expression for the electronic
diffusion coefficient $D=v_t l_e/3=29\rm\ nm^2/fs$, where  $l_e = v_t
\tau_{e-ph}$ is the electronic mean free path. It is interesting to
note that at $t=0.5$~ps the excited 
electrons have already  reached an average distance of 120~nm, about
ten times larger than the optical penetration depth. This is in
agreement with a value of  100~nm used to describe the initial spatial
distribution of excited electrons in the two-temperature
model.\cite{Hohlfeld00} Such a model  does not describe the
thermalization of the electron gas and starts to be valid only after
$t \ge 0.5$~ps.  Thus, one has to add $\langle z(t=0.5{\rm\
ps})\rangle$ to the optical penetration depth.

To conclude,  we have presented a theoretical model including both
transport and electron-phonon scattering which is able to  reproduce
the temperature dependence of the relaxation time. It is shown that
this variation is due to a drastic change of the excited-electron
motion, which is ballistic at low temperature and gets diffusive at
room temperature. Note that the correct order of magnitude of the
temperature effect is obtained without using free parameters: the
electron-phonon scattering rate was directly taken from experimental
data.  Furthermore, we also observe at low energy a delayed rise in
the 2PPE intensity and an increase due to temperature in agreement
with the experimental results.  As a final remark, we have provided a
very efficient method to describe the excited-electron dynamics in the
short-time regime, where electron-phonon energy transfer is negligible
(below $\approx 0.5$~ps). The results obtained here indicate that it
is promising to use our extended model to study other problems, in
particular the dynamics of excited electrons involving transport
effects  in the regime where electron-lattice energy transfer  becomes
relevant. So far, the only method available for this purpose is the
two-temperature model, which is not always reliable.\cite{Groeneveld}
The present theory is also suitable for the study of thin films, which
offer the possibility to probe the effect of transport. Our paper
shows how new information about the dynamics of excited electrons in
solids can be extracted from 2PPE.

We would like to thank H.~Petek, M.~Aeschlimann,  E.~Matthias, and
H.~C.~Siegmann for interesting discussions and helpful comments.
Financial support by Deutsche Forschungsgemeinschaft, Sfb 290 and 450,
is gratefully acknowledged.

\end{document}